\documentstyle[preprint,tighten,eqsecnum,aps,floats,epsfig]{revtex}

\begin{document}
\draft
\title{
$\Lambda$-parameter of lattice QCD with the overlap-Dirac operator}
\author{C. Alexandrou$^a$, H. Panagopoulos$^a$, E. Vicari$^b$}
\address{
$^a$Department of Natural Sciences, University of Cyprus,
P.O.Box 20537, Nicosia CY-1678, Cyprus.}
\address{
$^b$Dipartimento di Fisica dell'Universit\`a 
and I.N.F.N., 
Via Buonarroti 2, I-56127 Pisa, Italy.}

\date{\today}

\maketitle

\begin{abstract}
We compute the ratio  $\Lambda_L/\Lambda_{\overline{MS}}$
between the scale parameter $\Lambda_L$,
associated with a lattice formulation of QCD  
using the overlap-Dirac operator, and $\Lambda_{\overline{MS}}$
of the $\overline{\rm MS}$ renormalization scheme.
To this end, 
the necessary one-loop relation 
between the lattice coupling $g_0$ and the coupling renormalized
in the $\overline{{\rm MS}}$ scheme 
is calculated, using the lattice background field technique.


\end{abstract}


\section{Introduction}
\label{introduction}

The overlap-Dirac operator~\cite{Neuberger-98} that has been derived 
from the overlap formulation of chiral fermions on the lattice~\cite{N-N-95}
provides a lattice regularization of massless QCD, i.e. of chiral fermions
coupled vectorially to a gauge field,  without the need of fine tuning. 
It preserves chiral symmetry without fermion doubling, circumventing
the Nielsen-Ninomiya theorem~\cite{N-N-81}.
The overlap-Dirac operator $D_o$ of a  massless fermion
can be written as ($a$ is the lattice spacing)
\begin{equation}
D_o = {1\over a} \left[  1 + \gamma_5 O \right],
\end{equation}
where $O$ depends on the link variables, and is Hermitian with
eigenvalues $\pm 1$.
The simplest such example is given by
the Neuberger-Dirac operator~\cite{Neuberger-98}
\begin{eqnarray}
D_{\rm N} &=& {1\over a} \,c \left[  1 + X (X^\dagger X)^{-1/2} \right],\\
X &=& D_{\rm W} - {1\over a}\rho,
\label{Nop}
\end{eqnarray}
where $D_{\rm W}$ is the Wilson-Dirac operator (with the Wilson 
parameter $r$ set to its standard value, $r=1$)
\begin{equation}
D_{\rm W} = {1\over 2} \left[ \gamma_\mu \left( \nabla_\mu^*+\nabla_\mu\right)
 - a\nabla_\mu^*\nabla_\mu \right],
\end{equation}
\begin{equation}
\nabla_\mu\psi(x) = {1\over a} \left[ U(x,\mu) \psi(x + a\hat{\mu})
- \psi(x)\right],
\end{equation}
and $\rho$ is a real parameter subject to the constraint 
$0< \rho < 2$.\footnote{
Nonperturbatively one expects
$-m_c< \rho < 2$, where $m_c$ is the {\it negative} critical mass associated
with the Wilson-Dirac operator.} 
The normalization constant $c = \rho$ 
can be absorbed in the definition of the fermion fields.
Unlike the Wilson-Dirac operator $D_{\rm W}$, 
$D_N$ is not analytic in the link variables 
when the operator $X$ in Eq.~(\ref{Nop}) has a zero eigenvalue.
As discussed in Ref.~\cite{Neuberger-99},
this is the price to pay for putting
strictly massless fermions on the lattice. 
However, such a lack of analiticity is expected to be harmless in the 
continuum limit~\cite{Neuberger-99,H-J-L-99}.

$D_{\rm N}$ satisfies 
the Ginsparg-Wilson relation~\cite{G-W-82}
\begin{equation}
\gamma_5 D + D\gamma_5 = a D\gamma_5 D,
\label{GWrel}
\end{equation}
which protects the quark masses from 
additive renormalization~\cite{Neuberger-98,Hasenfratz-98}.
The Ginsparg-Wilson relation allows us to write,
at finite lattice spacing, relations that are 
essentially equivalent to those holding in the low-energy
phenomenology associated with chiral symmetry
(see e.g. Refs.~\cite{Chandrasekharan-99,K-Y-99-2}).
It indeed 
implies the existence of an exact chiral symmetry of the lattice action
under the transformation~\cite{Luscher-98}
\begin{equation}
\delta\psi(x) = \gamma_5 ( 1 - \case{1}{2}a D)\psi(x),
\qquad\qquad \delta\bar{\psi}(x) = \bar{\psi}(x)(1-\case{1}{2}a D)\gamma_5,
\label{latchsym}
\end{equation}
thus leading to chiral Ward identities that ensure the
non-renormalization of vector and flavor non-singlet axial vector
currents, and the absence of mixing among operators in different chiral
representations.
The axial anomaly then arises from the non-invariance of the
fermion integral measure under flavour-singlet chiral 
transformations
\cite{Luscher-98,Luscher-99,Chiu-99,Fujikawa-99,Suzuki-99,Adams-98}.
It is also worth mentioning
that lattice gauge theories with Ginsparg-Wilson
fermions have been proved to be  renormalizable to all orders 
of perturbation theory~\cite{R-R-99}.

The important point is that
lattice Dirac operators satisfying Eq.~(\ref{GWrel}) 
are not affected by the Nielsen-Ninomiya theorem~\cite{N-N-81},
thus they need not suffer from fermion doubling.
A lattice formulation of QCD satisfying the Ginsparg-Wilson relation
would overcome the complications of the standard
approach (e.g. Wilson fermions), where chiral symmetry is
violated at the scale of the lattice spacing.
As a consequence of chiral symmetry, 
leading scaling corrections are $O(a^2)$, as opposed to $O(a)$ of the chiral
symmetry violating case.

Indeed, $D_{\rm N}$ avoids fermion doubling. However,
its locality properties in the presence of a gauge field are not obvious.
$D_{\rm N}$ is not strictly local, and locality should be recovered 
only in a more general sense, i.e. allowing 
an exponential decay of the kernel of $D_{\rm N}$ with a rate which scales
with the lattice spacing and not with the physical quantities.
In Ref.~\cite{H-J-L-99} the locality of $D_{\rm N}$ has been proved
for sufficiently smooth gauge fields. Moreover numerical evidence has been presented for 
typical gauge fields in present-day simulations.
Thus, the Neuberger-Dirac operator seems to have all the right
properties that a lattice Dirac operator should have in order to describe massless
quarks. A major open question seems to be
its practical implementation, since $D_{\rm N}$  appears much more complicated
than the usual Wilson-like Dirac operators.
In this respect some progress has been achieved (see, e.g., 
Refs.~\cite{Neuberger-98-2,H-J-L-99,E-H-N-99,E-H-N-99-2,Neuberger-98-3,Neuberger-99,E-H-N-99-3,G-H-R-99,H-J-L-99,UKQCD-99,L-D-L-Z-99}),
and simulations may become feasible in the near future.

In this paper we calculate the ratio between the $\Lambda$-parameter 
associated with a lattice formulation of QCD  
using the overlap-Dirac operator 
($\Lambda_L$) and that of
the $\overline{\rm MS}$ renormalization scheme ($\Lambda_{\overline{\rm MS}}$).
In order to have a  complete discretization of QCD, 
we will consider the Wilson formulation for the pure gauge 
part of the theory. Actually, the part of the calculation involving fermions is
independent of the regularization chosen for the gluonic part.
We recall that 
the  ratio of any renormalization group invariant
quantity to the appropriate power of $\Lambda_L$ approaches
a constant in the continuum limit $g_0\rightarrow 0$.
Indeed $\Lambda_L$ is a particular solution of the renormalization group
 equation 
\begin{equation}
\left( -a{\partial\over \partial a} + \beta_L(g_0)
{\partial\over \partial g_0} \right)\Lambda_L = O(a),
\end{equation}
i.e.
\begin{equation}
a \Lambda_L = \exp\left[ - \int^{g_0} {dg\over \beta_L(g)}\right] 
= \exp \left( -{1\over 2b_0g_0^2}\right)
(b_0g_0^2)^{-{b_1/2b_0}}
\left[ 1 + O\left(g_0^2\right)\right] ,
\end{equation}
where $\beta_L(g_0)$ is the lattice $\beta$-function,
and $b_0, b_1$ the first two coefficients of its perturbative expansion:
\begin{eqnarray}
&&b_0 = {1\over (4\pi)^2} 
\left({11\over 3}N-{2\over 3}N_f\right),\\
&&b_1= {1\over (4\pi)^4} \left[{34\over 3}N^2 - N_f \left(
{13\over 3}N- {1\over N}\right)\right]
\end{eqnarray}
in $SU(N)$ gauge theory with $N_f$ fermion species.
The calculation of the ratio $\Lambda_L/\Lambda_{\overline{MS}}$
requires a one-loop perturbative calculation  
on the lattice. As it turns out,
the necessary computation is much more cumbersome than in the case of
Wilson fermions, due to the more complicate structure of the 
Neuberger-Dirac operator.

\section{Formulation of the problem}
\label{sec2}

The lattice regularization of QCD we consider is described by the action
\begin{equation}
S_L =  {1\over g_0^2} \sum_{x,\mu,\nu}
{\rm Tr}\left[ 1 - U_{\mu\nu}(x) \right]  +
 \sum_{i=1}^{N_f} \sum_{x,y} \bar{\psi}_i(x)D_{\rm N}(x,y)\psi_i(y).
\label{latact}
\end{equation}
where $U_{\mu\nu}(x)$ is the usual product of link variables
$U_{\mu}(x)$ along the perimeter of a plaquette
originating at $x$ in the positive $\mu$-$\nu$ directions,
and $N_f$ is the number of massless fermions considered.
We recall that $\Lambda_L$ is independent of the fermionic masses,
so the results of our calculation will also hold for the massive cases.
\footnote{
In order to describe massive fermions one may write
the overlap-Dirac operator in the form~\cite{Neuberger-98-4,E-H-N-99-2}
$ D_o \propto 1 + \mu  + (1-\mu) \gamma_5 O .$ 
Values $0\leq \mu \leq 1$
describe fermions with mass $m_f$, $0\le m_f < \infty$.
For small $\mu$, $m_f$ is proportional to $\mu$.}

In order to evaluate the ratio $\Lambda_{\rm L}/\Lambda_{\overline{\rm MS}}$
we need to calculate, at one loop order, the relation
between the lattice coupling $g_0$ and the renormalized 
coupling $g$ of the $\overline{\rm MS}$ scheme:
\begin{equation}
g_0 = Z_g(g_0,a\mu) g,
\label{grenorm}
\end{equation}
where $\mu$ indicates a renormalization scale.
Writing
\begin{equation}
Z_g(g_0,x)^2  = 1 + g_0^2\left( 2 b_0 \ln x + l_0 \right) + O(g_0^4), 
\label{zgzg}
\end{equation}
one has
\begin{equation}
l_{0} = 2 b_0\ln \left( \Lambda_L/\Lambda_{\overline{\rm MS}}\right).
\label{l0}
\end{equation}

The computation of $Z_g$ is easier in the background field gauge.
In fact, this renormalization constant
has a simple relationship with the
background field renormalization constant
$Z_A$~\cite{Abbott-81},
\begin{equation}
Z_A(g_0,a\mu)   Z_g(g_0, a\mu)^2 = 1.
\label{eq:zeqza}
\end{equation}
As a consequence of this relation,
in order to calculate $Z_g$
one only needs to calculate the one-loop self-energy
of the background field.

The $\overline{\rm MS}$ renormalized one-particle irreducible
two-point function of the background field is given by
\begin{equation}
\Gamma^{AA}_R(p)^{ab}_{\mu\nu} =
-\delta^{ab}\left( \delta_{\mu\nu}p^2 - p_\mu p_\nu\right)
\left( 1 - \sum_{i=1} g^{2i} \nu^{(i)}_R(p/\mu)\right)/g^2, 
\end{equation}
where 
\begin{equation}
\nu^{(1)}_R(p/\mu,\lambda)=
{N\over 16\pi^2} \left[  
-{11\over 3}\ln {p^2\over \mu^2} +{205\over 36}
+{3\over 2\lambda} + {1\over 4 \lambda^2}\right] +
{N_f\over 16\pi^2}\left[ {2\over 3}\ln{p^2\over \mu^2}
- {10\over 9}\right]
\label{nur}
\end{equation}
and $\lambda$ is the gauge parameter.
On the lattice one writes
\begin{equation}
\sum_\mu \Gamma^{AA}_L(p)^{ab}_{\mu\mu} =
-\delta^{ab}3\widehat{p}^2 
\left[ 1 - \sum_{i=1}g_0^{2i}\nu^{(i)}(ap)\right]/g_0^2.
\end{equation}
The bare and renormalized functions
are related by
\begin{equation}
\left[ 1 - \sum_{i=1} g^{2i} \nu^{(i)}_R(p/\mu) \right]
= Z_A \left[ 1 - \sum_{i=1}g_0^{2i} \nu^{(i)}(ap) \right].
\label{eq:selfs}
\end{equation}
Therefore, using relation (\ref{eq:zeqza}),
\begin{equation}
Z_g^2 = {1 - \sum_{i=1} g_0^{2i}\nu^{(i)}(ap)\over 
1 - \sum_{i=1}g^{2i}\nu^{(i)}_R(p/\mu)}.
\label{zgzgzg}
\end{equation} 
So, in order to evaluate $l_0$ we have to perform the
one-loop calculation of the function
$\nu^{(1)}(ap)$ in the background field gauge
as formulated on the lattice~\cite{D-G-81,L-W-95}. 
We mention that 
the background field technique on the lattice
has been recently employed
for the calculations of the third coefficient of the lattice
$\beta$-function for Wilson-like lattice 
formulations of QCD~\cite{L-W-95-2,C-F-P-V-98,Haris-99}.
We refer to the above cited references for the relevant
formulae.

In order to perform the lattice perturbative calculation of
$\nu^{(1)}(ap)$ we must formally expand $D_{\rm N}$ in powers of $g_0$. 
The weak coupling expansion of $D_{\rm N}$ has been 
already discussed in Ref.~\cite{K-Y-99}. We list here the relevant
formulae for our one-loop calculation.

Let us first write down the weak coupling expansion of
the Wilson-Dirac operator $D_{\rm W}$. This will be useful for
constructing the relevant vertices of $D_{\rm N}\,$.
We write
\begin{equation}
X(q,p) = D_{\rm W}(q,p) - {1\over a}\rho = 
X_0(p)(2\pi)^4\delta^4(q-p) + X_1(q,p) + X_2(q,p) + O(g_0^3),
\label{aqp}
\end{equation}
where
\begin{equation}
X_0(p)= {i\over a} \sum_\mu \gamma_\mu \sin a p_\mu + {1\over a} \sum_\mu
(1-\cos a p_\mu) - {1\over a} \rho,
\label{a0}
\end{equation}
\begin{eqnarray}
X_1(q,p) &=& g_0 \int d^4 k \delta(q-p-k) A_\mu(k) V_{1,\mu}(p+k/2),\label{a1}\\
V_{1,\mu}(q) &=&
i\gamma_\mu \cos aq_\mu + \sin a q_\mu,\nonumber
\end{eqnarray}
\begin{eqnarray}
X_2(q,p) &=& {g_0^2\over 2} \int {d^4 k_1 \, d^4 k_2\over (2\pi)^4}
 \delta(q-p-k_1-k_2) A_\mu(k_1)A_\mu(k_2) V_{2,\mu}(p+k_1/2+k_2/2),\label{a2}
\\
V_{2,\mu}(q) &=&
-i\gamma_\mu a\sin aq_\mu + a \cos a q_\mu.\nonumber
\end{eqnarray}

Let us write the Fourier transform of the Neuberger-Dirac operator in the form
\begin{equation}
D_{\rm N}(q,p) = D_0(p) (2\pi)^4\delta^4(q-p) + \Sigma(q,p).
\end{equation}
$D_0(p)$ is the tree level inverse propagator:
\begin{equation}
D_0^{-1}(p) = {-i\sum_\mu \gamma_\mu \sin ap_\mu \over 2 \left[ \omega(p) + b(p)\right] }
+ {a\over 2},
\label{d0}
\end{equation}
where
\begin{eqnarray}
\omega(p) &=& {1\over a} \left( \sum_\mu \sin^2 ap_\mu + \bigl[ 
\sum_\mu (1-\cos ap_\mu ) - \rho \bigr]^2 \right)^{1/2}, \\
b(p)&=& {1\over a} \sum_\mu (1-\cos ap_\mu) - {1\over a} \rho.
\end{eqnarray}
One can easily check that for $0<\rho<2$ the propagator has only one pole,
at $p=0$.
The function $\Sigma(q,p)$ can be expanded in powers of $g_0$ as~\cite{K-Y-99}
\begin{eqnarray}
a\Sigma(q,p) = &&
{1\over \omega(p) + \omega(q)}
\left[X_1(q,p) - {1\over \omega(p)\omega(q)} X_0(p) X^\dagger_1(p,q) X_0(q)\right] \nonumber\\
&&+
{1\over \omega(p) + \omega(q)}
\left[X_2(q,p) - {1\over \omega(p)\omega(q)} X_0(p) X^\dagger_2(p,q) X_0(q)\right] \nonumber\\
&&+
\int {d^4 k\over (2\pi)^4}
{1\over \omega(p) + \omega(q)}
{1\over \omega(p) + \omega(k)}
{1\over \omega(q) + \omega(k)}\times \nonumber \\
&&\;\;\Biggl[
-X_0(p)X_1^\dagger(p,k)X_1(k,q) 
 -X_1(p,k)X_0^\dagger(k)X_1(k,q) -X_1(p,k)X_1^\dagger(k,q)X_0(q) \nonumber \\
&&\;\;+{\omega(p)+\omega(q)+\omega(k)\over \omega(p)\omega(q)\omega(k)}
X_0(p)X_1^\dagger(p,k)X_0(k)X_1^\dagger(k,q)X_0(q)\Biggr] + ...
\label{vertices}
\end{eqnarray}
From $\Sigma(q,p)$ one can read off the vertices necessary for the
one loop calculation of $\nu^{(1)}(ap)$.

\section{Results and discussion}
\label{sec3}

Two diagrams containing fermions contribute to $\nu^{(1)}(p)$, shown
in Figure 1. Given that the 4-point vertex contains a part with an internal
momentum ($k$ in Eq.~(\ref{vertices})), the corresponding part of the second 
diagram actually has the same connectivity as the first diagram.

\begin{figure}[tb]
\hfil\mbox{\epsfysize=5cm\epsfxsize=10cm\epsffile{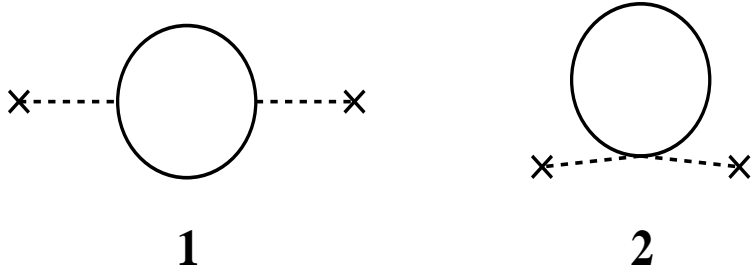}}\hfil
\caption{
Diagrams with fermionic lines contributing
to the one-loop function $\nu^{(1)}(ap)$.
Dashed lines ending on a cross represent background gluons.
Solid lines represent fermions.
}
\label{fig1}
\end{figure}

The algebra involving lattice quantities was performed using a
symbolic manipulation package which we have developed in Mathematica. 
For the purposes of the present work, this package was
augmented to include the propagator and vertices of the overlap action.

\medskip
The one-loop amplitude $\nu^{(1)}(ap)$ can be written as:
\begin{equation}
\nu^{(1)}(ap) = \nu^{(1)}(ap)|_{N_f{=}0} + N_f \cdot \sum_i \nu_i(ap)
\end{equation}
(the index $i$ runs over the two diagrams shown in Figure 1), where,
\begin{equation}
\widehat{ap}^2 \nu_i(ap) = d_{0,i} + a^2 p^2 \left\{ d_{1,i} +
d_{2,i}\, {\ln a^2 p^2 \over (4 \pi)^2} \right \} + O((ap)^4) 
\end{equation}
($\widehat{p}^2 = 4 \sum_\mu \sin^2(p_\mu/2)$). The coefficients 
$d_{n,i}$ depend on $\rho$, but not on $N$ or  $N_f\,$. 
Note that the diagrams 
in Figure 1 do not involve the gluonic propagator, so they are independent of 
the choice of regularization for the pure gluonic part of the action.

To extract the $p$-dependence, we first isolate the 
divergent terms; these are responsible for the 
logarithms. There are only a few such terms, and in the pure
gluonic case their values are well known. We can use
these values also in diagrams with fermions, applying successive
subtractions of the type: 
\begin{equation}
{1\over \overline{q}^2} = 
{1\over \widehat{q}^2} + \left( {1\over \overline{q}^2} -
{1\over \widehat{q}^2}\right)
\end{equation}
where $\overline{q}^2$ is the 
inverse fermionic propagator.
All remaining terms now contain no divergences, and can be evaluated by Taylor
expansion in $ap$.

At this stage, one is left with expressions which no longer contain
$p$ and must be numerically integrated over the loop momentum.
Given the complicated form of the overlap vertices, these expressions turn out 
to be quite lengthy, containing a few thousand terms. 

\begin{table}[tbp]

\begin{minipage}{3cm}
\hfill
\end{minipage}
\begin{minipage}{10cm}
\caption{Coefficients $d_{1,i}(\rho)$.}
\label{tab1}
\begin{tabular}{ccc}
{$\rho$}&$d_{1,1}$ &$d_{1,2}$\\
\tableline \hline
0.2  & $-$0.00222139 &  0.01803836  \\
0.3  & $-$0.00418268 &  0.01753315  \\
0.4  & $-$0.00563107 &  0.01732216  \\
0.5  & $-$0.00681650 &  0.01727870  \\
0.6  & $-$0.00785477 &  0.01736058  \\
0.7  & $-$0.00881237 &  0.01755678  \\
0.8  & $-$0.00973458 &  0.01787211  \\
0.9  & $-$0.01065738 &  0.01832255  \\
1.0  & $-$0.01161402 &  0.01893460  \\
1.1  & $-$0.01263969 &  0.01974719  \\
1.2  & $-$0.01377630 &  0.02081600  \\
1.3  & $-$0.01507869 &  0.02222125  \\
1.4  & $-$0.01662489 &  0.02408181  \\
1.5  & $-$0.01853492 &  0.02658162  \\
1.6  & $-$0.02101054 &  0.03002396  \\
1.7  & $-$0.02443278 &  0.03495886  \\
1.8  & $-$0.02966205 &  0.04255372  \\
\end{tabular}
\end{minipage}
\end{table}

The integration is done in momentum space over finite lattices;
an extrapolation to infinite size is then performed, in the manner of 
Ref.~\cite{C-F-P-V-98}. We evaluated the integrals for a range of values of 
the parameter $\rho$, as presented in Table 1. 
For all values of $\rho$ tabulated, 
lattice sizes $L\le 128$ were sufficient to yield answers with at least 5
significant digits (the uncertainty coming from a systematic error in the 
extrapolation, which can be estimated quite accurately). As the endpoints of 
the perturbative domain of $\rho$ are approached ($\rho \to 0,\ \rho \to 2$), 
increasingly 
larger lattices are required for similar accuracy; this is, of course, a
reflection of the divergences in the propagator at these endpoints. 
As shown in Figure 2, the value of the 
$\Lambda_L$-parameter varies also in a more pronounced manner near the 
endpoints, where it is expected to diverge.

\begin{figure}[tb]
\mbox{\epsfysize=12cm\epsfxsize=16cm\epsffile{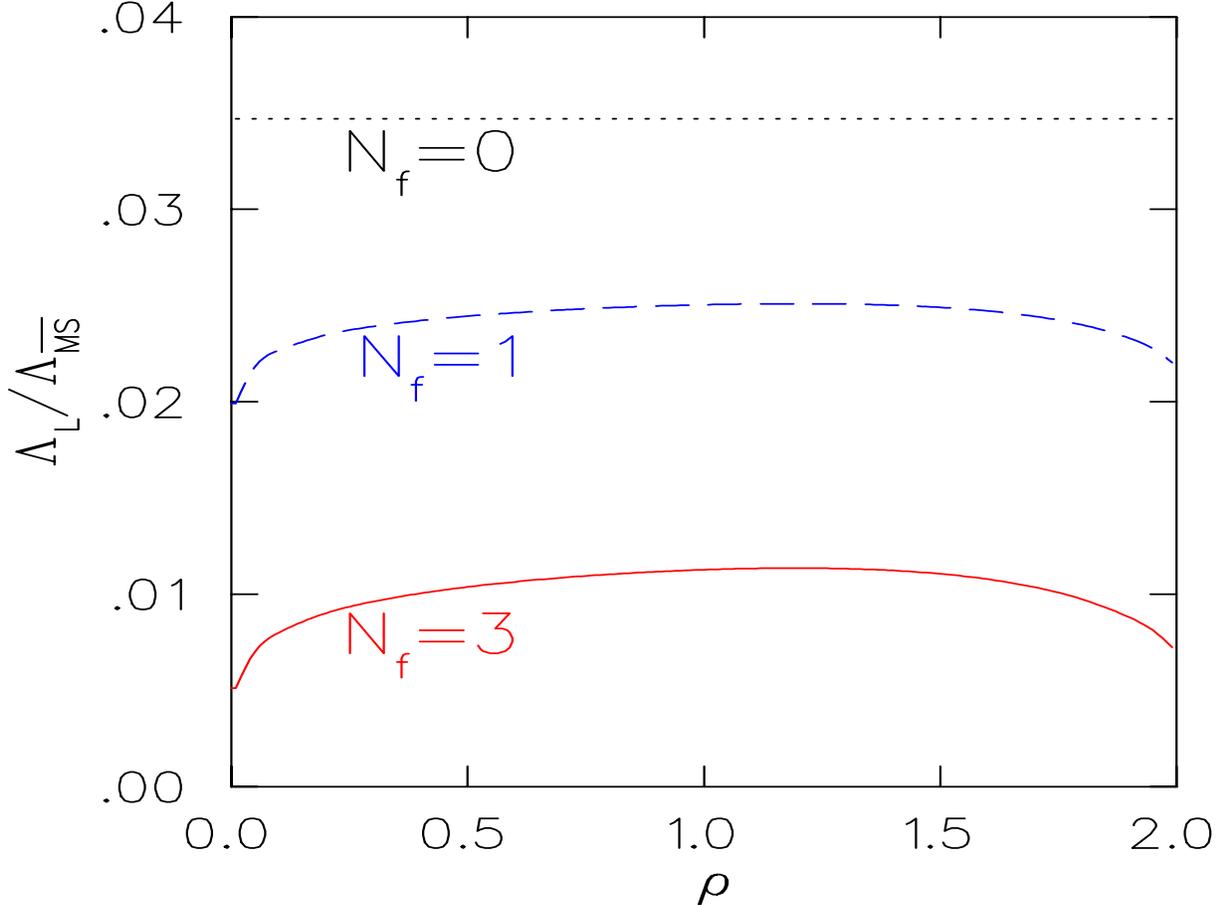}}
\caption{
$\Lambda_L/\Lambda_{\overline{\rm MS}}$ in $SU(3)$, as a function of the 
parameter $\rho\,$. $N_f = 0$ (dotted line), $N_f = 1$ (dashed line), $N_f = 3$ 
(solid line).
}
\label{fig2}
\end{figure}

The coefficients $d_{n,i}$ obey two constraints, which we have used
as verifications of our procedure:
\begin{itemize}
\item[i)] Gauge invariance requires $\sum_i d_{0,i} = 0$.
One may check this property both on the algebraic expressions and on 
the numerical results.
Substituting the numerical results for the coefficients, we find zero within
the error estimates, for all values of $\rho\,$; this serves also to verify our
estimation of systematic errors.

\item[ii)] The coefficients $d_{2,i}$ must coincide with those
of the continuum. We checked that this is so:
$d_{2,1} = {2/ 3}\, , \ d_{2,2} = 0\, .$

\end{itemize}

\medskip
Thus, our result takes the form
\begin{equation}
\nu^{(1)}(ap)\mid_{\lambda_0=1} = \nu^{(1)}(ap)\mid_{N_f=0,\;\lambda_0=1} + 
N_f \left[{1\over 24\pi^2}\ln (a^2p^2) + k_{f}(\rho) \right],
\label{nu1}
\end{equation}
where 
\begin{equation}
\nu^{(1)}(ap)\mid_{N_f=0,\;\lambda_0=1} =
-{11 N\over 48\pi^2} \ln (a^2p^2) - {1\over 8 N} + 0.21709849 N
\end{equation}
is the pure gauge part~\cite{H-H-81,G-K-82,L-W-95-2}, 
and $k_{f}(\rho) = d_{1,1}
+ d_{1,2}\,$ is presented in Table 1. We note again that the choice 
of the pure gluonic part of the action affects the first, but not the second, 
summand in Eq.~(\ref{nu1}).

From Eqs.~(\ref{zgzg}, \ref{nur}, \ref{zgzgzg}), we find
\begin{equation}
l_0 = {1\over 8 N} - 0.16995599 N
+ N_f\left[ - {5\over 72 \pi^2} - k_{f}(\rho)\right].
\end{equation}

The values of $\Lambda_L/\Lambda_{\overline{\rm MS}}$ follow immediately from 
Eq.~(\ref{l0}). Some particular cases of interest are ($N=3$):

\begin{eqnarray}
&&\Lambda_L/\Lambda_{\overline{\rm MS}}\bigr|_{N_f{=}0} = 0.034711\\
&&\Lambda_L/\Lambda_{\overline{\rm MS}}\bigr|_{N_f{=}1,\ \rho{=}1} = 0.025042\\
&&\Lambda_L/\Lambda_{\overline{\rm MS}}\bigr|_{N_f{=}3,\ \rho{=}1} = 0.011273
\end{eqnarray}

For comparison we report 
the corresponding values of the ratio $\Lambda_L/\Lambda_{\overline{\rm MS}}$
for the standard Wilson formulation of fermions~\cite{K-N-S-81}
\begin{eqnarray}
&&\Lambda_L/\Lambda_{\overline{\rm MS}}\bigr|_{N_f{=}1,\ r{=}1} = 0.029412\\
&&\Lambda_L/\Lambda_{\overline{\rm MS}}\bigr|_{N_f{=}3,\ r{=}1} = 0.019618.
\end{eqnarray}

\bigskip\bigskip
\noindent
{\bf Acknowledgements:} H. P. would like to acknowledge the warm
hospitality extended to him by the Theory Group in Pisa during various
stages of this work.


\end{document}